\documentclass[11pt]{article}
\usepackage{graphicx}
\usepackage{caption}
\usepackage{authblk}
\usepackage{amsthm}
\usepackage[letterpaper, left=1in, right=1in, bottom=1in, top=1in]{geometry}
\usepackage{color}
\usepackage{amsmath}
\usepackage{amssymb}
\usepackage{mathtools}
\usepackage{hyperref}
\usepackage[T1]{fontenc}
\usepackage{diagbox}
\usepackage{url}
\usepackage{hyperref}
\usepackage{cleveref}
\usepackage{physics}
\usepackage{array}
\usepackage{algorithm}
\usepackage{algpseudocode}

\newcommand{\lrceil}[1]{\lceil #1 \rceil}
\newcommand{\ol}[1]{\overline{#1}}
\newcommand{\mtt}[1]{\mathtt{#1}}
\newcommand{\mrm}[1]{\mathrm{#1}}

\newcommand{\mcal}[1]{\mathcal{#1}}

\newcommand{\SubSeq}{\mathit{SubSeq}}
\newcommand{\SubSeqf}{\mathit{SubSeq^{\le f}}}
\newcommand{\lcsuf}{\mathsf{lcsuf}}

\newcommand{\SegSub}[1]{\mathit{SubSeq}^{\le #1}}
\newcommand{\SegSuf}[1]{\mathit{SufSeq}^{\le #1}}
\newcommand{\Segb}[1]{\ol{\mathit{SubSeq}}_{\ge #1}}
\newcommand{\Segf}[1]{\ol{\mathit{SufSeq}}_{\ge #1}}
\newcommand{\maxlen}{\mathrm{mxl}}
\newcommand{\lpf}{\mathit{lpf}}
\newcommand{\lsf}{\mathit{lsf}}
\newcommand{\llpf}{\mathit{llpf}}

\newcommand{\tta}{\texttt{a}}
\newcommand{\ttb}{\texttt{b}}
\newcommand{\ttc}{\texttt{c}}

\newcommand{\SegSPM}{\textsf{SegE}}
\newcommand{\SegSEM}{\textsf{MinSegE}}
\newcommand{\SegLCS}{\textsf{SegLCS}}
\newcommand{\GenSegLCS}{\textsf{IndSegLCS}}
\newcommand{\LowestRowNum}{\mathsf{MaxVIdx}}

\newcommand{\LCSk}{\mathsf{slcs}}

\newtheorem{theorem}{Theorem}
\newtheorem{corollary}{Corollary}
\newtheorem{lemma}{Lemma}

\newtheorem{conjecture}{Conjecture}
\newtheorem{example}{Example}
\theoremstyle{definition}

\newtheorem{problem}{Problem}

\newlength{\infwidth}
\newcolumntype{C}{wc{0.90\infwidth}}

\begin{document}
\title{Subsequence Matching and LCS\texorpdfstring{\\}{} with Segment Number Constraints}
\author[1]{Yuki~Yonemoto}
\author[2]{Takuya~Mieno}
\author[1]{Shunsuke~Inenaga}
\author[3]{Ryo~Yoshinaka}
\author[3]{Ayumi~Shinohara}

\affil[1]{Kyushu University, Japan}
\affil[2]{University of Electro-Communications, Japan}
\affil[3]{Tohoku University, Sendai, Japan}

\date{}
\maketitle

\begin{abstract}
  The \emph{longest common subsequence} (\emph{LCS}) is a fundamental problem in string processing
  which has numerous algorithmic studies, extensions, and applications.
  A sequence $u_1, \ldots, u_f$ of $f$ strings 
  is said to be an ($f$-)segmentation of a string $P$ if $P = u_1 \cdots u_f$.
  Li et al. [BIBM 2022] proposed a new variant of the LCS problem
  for given strings $T_1, T_2$ and an integer $f$,
  which we hereby call the segmental LCS problem ({\SegLCS}),
  of finding (the length of) a longest string $P$ 
  that has an $f$-segmentation which can be embedded into both $T_1$ and $T_2$.
  Li et al. [IJTCS-FAW 2024] gave a dynamic programming solution that solves {\SegLCS} in  $O(fn_1n_2)$ time
  with $O(fn_1 + n_2)$ space, where $n_1 = |T_1|$, $n_2 = |T_2|$, and $n_1 \le n_2$.
  Recently, Banerjee et al. [ESA 2024] presented an algorithm which, for a constant $f \geq 3$, 
  solves {\SegLCS} in $\tilde{O}((n_1n_2)^{1-(1/3)^{f-2}})$ time\footnote{$\tilde{O}(\cdot)$ suppresses polylogarithmic factors.}.
  In this paper, we deal with {\SegLCS} as well as the problem of segmental subsequence pattern matching,
  {\SegSPM}, that asks to determine whether 
  a pattern $P$ of length $m$ has an $f$-segmentation that can be embedded into a text $T$ of length $n$.
  When $f = 1$, this is equivalent to substring matching, and when $f = |P|$,
  this is equivalent to subsequence matching.
  Our focus in this article is the case of general values of $f$, and our main contributions are threefold:
  \begin{enumerate}
    \item[(1)] $O((mn)^{1-\epsilon})$-time conditional lower bound for {\SegSPM} under the strong exponential-time hypothesis (SETH), for any constant $\epsilon > 0$.
    \item[(2)] $O(mn)$-time algorithm for {\SegSPM}.
    \item[(3)] $O(fn_2(n_1 - \ell+1))$-time algorithm for {\SegLCS} where $\ell$ is the solution length.
  \end{enumerate}
\end{abstract}

\section{Introduction}\label{sec:intro}

The \emph{longest common subsequence} (\emph{LCS}) is a fundamental problem in string processing
which has numerous algorithmic studies, extensions, and applications.
Li et al.~\cite{LiZJFC22} proposed a new variant of the LCS problem for given string $T_1, T_2$ and integer $f$,
which we hereby call the \emph{segmental LCS} problem ({\SegLCS}).
A sequence $u_1, \ldots, u_f$ of $f$ strings 
is said to be an ($f$-)segmentation of a string $P$ if $P = u_1 \cdots u_f$.
{\SegLCS} asks to find (the length of) a longest string $P$ 
that has an $f$-segmentation which can be embedded into both $T_1$ and $T_2$.
Throughout this paper,
we assume $n_1 = |T_1| \le n_2 = |T_2|$ without loss of generality.
Li et al.~\cite{LiJCFZ24} gave a dynamic programming solution that solves {\SegLCS} in $O(fn_1n_2)$ time
with $O(fn_1 + n_2)$ space.
Recently, Banerjee et al.~\cite{BanerjeeGT24} presented an algorithm which, for a constant $f \geq 3$, 
solves {\SegLCS} in $\tilde{O}((n_1n_2)^{1-(1/3)^{f-2}})$ time.
The framework of Banerjee et al.~\cite{BanerjeeGT24} is general enough to solve other related problems, 
including the segmental version of the \emph{Episode Subsequence Matching} 
in $\tilde{O}((n_1n_2)^{1-(1/3)^{f-2}})$ time for a constant $f \geq 3$.

In this paper, we deal with the general case where $f$ is not necessarily a constant,
and our focus is to analyze the complexities of these problems.
Namely, we consider {\SegLCS} as well as the decision version of the segmental subsequence pattern matching,
{\SegSPM}, that asks to determine whether $P$ of length $m$ has an $f$-segmentation
that can be embedded into a text $T$ of length $n$.
When $f = 1$, this is equivalent to the standard substring pattern matching, and when $f = |P|$,
this is equivalent to the standard subsequence pattern matching, both of which admit linear-time solutions.

Our focus in this article is the case of general values of $f$.
The main contributions of this paper are threefold:
\begin{enumerate}
  \item[(1)] $O((mn)^{1-\epsilon})$-time conditional lower bound for {\SegSPM} (Problem~\ref{def:SegSPM}) under the strong exponential-time hypothesis (SETH), for any constant $\epsilon > 0$.
  \item[(2)] $O(mn)$-time algorithm for {\SegSPM} (Problem~\ref{def:SegSPM}).
  \item[(3)] $O(fn_2(n_1 - \ell+1))$-time algorithm for {\SegLCS} (Problem~\ref{def:SegLCS}) where $\ell$ denotes the length of the solution.
\end{enumerate}

Result (2) gives a matching upper bound with the conditional lower bound of Result (1) for {\SegSPM}.
Also, this algorithm works in $O(mn)$ time independently of the value of a given $f$,
and thus, it is faster than applying the $O(fmn)$-time {\SegLCS} solution by Li et al.~\cite{LiJCFZ24} to {\SegSPM}.
Our algorithm $O(f n_2 (n_1-\ell+1))$-time of Result (3) is based on
the approach by Banerjee et al.~\cite{BanerjeeGT24}
and Nakatsu et al.'s LCS algorithm~\cite{nakatsu1982longest},
which is at least as efficient as the $O(fn_1n_2)$-time solution by Li et al.~\cite{LiJCFZ24},
and can be faster when the solution is sufficiently long.
In particular, when $n_1 - \ell = O(1)$, our algorithm runs in $O(fn_2)$ time.

The rest of this paper is organized as follows:
We give basic notions and define our problems in Section~\ref{sec:preliminaries}.
In Section~\ref{sec:lowerbound} we present our lower bound for {\SegSPM},
and in Section~\ref{sec:SegSPM_algorithms} we give
an algorithm for {\SegSPM} running in $O(mn)$ time.
Section~\ref{sec:KMP} shows that {\SegSPM} can be solved in $O(m+n)$ time by a KMP-based algorithm if $f = 2$.
Section~\ref{sec:Nakatsu} is devoted for our $O(fn_2(n_1 - \ell+1))$-time algorithm for {\SegLCS}.
We conclude in Section~\ref{sec:conclusions}.

\section{Preliminaries} \label{sec:preliminaries}

Let $\Sigma$ be an alphabet.
An element of $\Sigma$ is called a character.
An element of $\Sigma^\star$ is called a string.
The empty string $\lambda$ is the string of length $0$.
For any strings $u$ and $v$, we denote by $u\cdot v$ the concatenation of the strings.
We often denote $uv = u\cdot v$ if there is no confusion.
For any string $u$ and a positive integer $e$, we define $u^0 = \lambda$ and $u^e = uu^{e-1}$.
For a string $T$, the length of $T$ is denoted by $|T|$.
If $T = xyz$ holds for some strings $x, y, z \in \Sigma^\star$, then $x$, $y$, and $z$ are called a prefix, a factor, a suffix of $T$, respectively.
For an integer $i$ with $1\le i \le |T|$, the $i$th character of $T$ is denoted by $T[i]$.
For integers $i, j$ with $1\le i\le j\le |T|$, the factor of $T$ starting at position $i$ and ending at position $j$ is denoted by $T[i.. j]$.
For two strings $S$ and $T$, we denote by $\lcsuf(S, T)$ the length of longest common suffix of $S$ and $T$.

A sequence $(u_1, \ldots, u_f)$ of $f$ strings is called an \emph{($f$-)segmentation} of its concatenation $u_1 \dots u_f$,
where each $u_i$ is said to be a \emph{segment} of the $f$-segmentation.
We say that a segmentation $(u_1, \ldots, u_f)$ of a string $P$ can be \emph{embedded} to another string $T$
if there exists a sequence $(p, g_1, \ldots, g_{f-1}, s)$ of $f+1$ (possibly empty) strings
such that $T = p \cdot u_1 \cdot g_1 \cdot u_2 \cdot g_2 \cdots g_{f-2} \cdot u_{f-1} \cdot g_{f-1} \cdot u_f \cdot s$.
Each $g_i$ is said to be a \emph{gap} in the embedding.
We say that $P$ is an \emph{$f$-segmental subsequence} of a string $T$ if
there exists an $f$-segmentation $(u_1, \ldots, u_f)$ of $P$ that can be embedded to $T$.
The set of $f$-segmental subsequences of $T$ is denoted by $\SegSub{f}(T)$.
Any element of $\SubSeq(T) = \bigcup_f \SegSub{f}(T)$ is called a \emph{subsequence}.

We consider the following decision problem: 
\begin{problem}[\SegSPM]\label{def:SegSPM}
  Given strings $T$ and $P$ and a positive integer $f$, decide whether $P \in \SegSub{f}(T)$.
\end{problem}
Note that since each gap $g_i$ in an embedding can be the empty string,
{\SegSPM} asks if $P$ has a segmentation with at most $f$ segments that can be embedded into $T$.

The following problem asks the minimum number of segments with which 
$P$ can be embedded into $T$:
\begin{problem}[\SegSEM]\label{def:SegNum}
  Given strings $T$ and $P$, 
  find the smallest positive integer $f$ such that $P \in \SegSub{f}(T)$ if it exists,
  and return $\mathit{nil}$ otherwise.
\end{problem}

We also consider the segmental version of the \emph{longest common subsequence} (\emph{LCS}) problem,
which we call {\SegLCS}\footnote{{\SegLCS} for $f$ is equivalent to the longest $f$-common sub-string problem (longest $f$-\textsf{CSS}) in~\cite{LiZJFC22} and
the $f$-longest common factor with gaps problem ($f$-\textsf{LCFg}) in~\cite{BanerjeeGT24}.}.
A pattern $P$ is said to be an \emph{$f$-segmental common subsequence} ($f$-\emph{SegCS}) of two strings $T_1$ and $T_2$
if $P$ has a segmentation $(u_1, \dots, u_f)$ of size $f$ that can be embedded to both $T_1$ and $T_2$.
We denote by $\LCSk(S, T, f)$ the length of a longest $f$-SegCS of $S$ and $T$.
\begin{problem}[\SegLCS]\label{def:SegLCS}
  Given strings $T_1, T_2$, and a positive integer $f$, compute $\LCSk(T_1, T_2, f)$.
\end{problem}

\section{$O((mn)^{1-\epsilon})$-time Conditional Lower Bound for {\SegSPM}} \label{sec:lowerbound}

In this section, we present a lower bound for {\SegSPM} such that 
there is no strongly sub-quadratic solution exists unless the famous \emph{strong exponential-time hypothesis} (\emph{SETH}) fails.

\begin{problem}[$k$-SAT problem] \label{prob:k-sat}
  Given a propositional logic formula of conjunctive normal form which has at most $k$ literals in each clause, decide whether there exists an interpretation that satisfies the input formula.
\end{problem}

\begin{conjecture}[The Strong Exponential-Time Hypothesis; SETH] \label{conj:seth}
  \leavevmode \\
  For any $\epsilon > 0$, there exists $k \ge 3$ such that the $k$-SAT problem cannot be solved in $O(2^{(1-\epsilon)n})$ time, where $n$ is the number of variables.
\end{conjecture}

Bille et al.~\cite{Bille2022} have shown a conditional lower bound on the complexity of \emph{Episode Matching}.
\begin{problem}[Episode Matching]
  Given two strings $T$ and $P$, compute a shortest factor $S$ of $T$ such that $P \in \SubSeq(S)$.
\end{problem}
\begin{theorem}[\cite{Bille2022}]\label{thm:EM_CLB}
  For any $\epsilon  > 0$ and any $\alpha \le 1$, Episode Matching on binary strings $T$ and $P$ with $|P| \in \Theta(|T|^\alpha)$ cannot be solved in $O((|T||P|)^{1-\epsilon})$ time, unless SETH is false.
\end{theorem}
This section proves a conditional lower bound on the complexity of {\SegSPM} by a reduction from Episode Matching.
We use Theorem~\ref{thm:EM_CLB} with $\alpha = 1$.
The proof by Bille et al.\ implies that only deciding whether a desired factor has a certain length is already as hard.
\begin{corollary}[\cite{Bille2022}]\label{cor:episodematching}
  For any $\epsilon  > 0$,  given binary strings $T$ and $P$ with $|P| \in \Theta(|T|)$ and an integer $h$,
  one cannot decide whether there is a factor $S$ of $T$ such that $|S| \le h$ and $P \in \SubSeq(S)$ in $O(|T|^{2-\epsilon})$ time, unless SETH is false.
\end{corollary} 

We show the following theorem using Corollary~\ref{cor:episodematching}.
\begin{theorem}\label{th:em2segspm}
  Neither {\SegSPM} nor {\SegSEM} over an alphabet of size three can be solved in $O((|T||P|)^{1-\epsilon})$ time for any $\epsilon > 0$, unless SETH fails.
\end{theorem}
\begin{proof}
  Since {\SegSEM} is an optimization version of {\SegSPM}, it suffices to show the hardness of {\SegSPM}.
  Let $T$ and $P$ be instance strings of Episode Matching over $\{0,1\}$ with $|T|=n$ and $|P|=m \in \Theta(n)$.
  Define two strings over $\{0,1,\$\}$ by
  \begin{align*}
    T' &= (\$ 0)^{2n-2} \$^2 T[1] \$^2 T[2] \$^2 \cdots \$^2 T[n] \$^2 (0\$)^{2n-2}\,,
    \\
    P' &= \$^{2n} P \$^{2n}
  \end{align*}
  (see Example~\ref{ex:em2segspm} below).
  Clearly $|T'|,|P'| \in \Theta(n)$. 
  By Corollary~\ref{cor:episodematching}, it is enough to show that $T$ has a factor of length $h$ subsuming $P$ as a subsequence if and only if $P' \in\SegSub{f}(T')$ where $f = 3n+m+h-4$.

  Suppose $P \in \SubSeq(T[i..j])$ where $j-i+1=h$.
  Then, $\$^2 P \$^2$ is an $m$-segmental subsequence of $S = \$^2 T[i] \$^2 \cdots \$^2 T[j] \$^2$ in $T'$, where the first and the last segments have length three and the others consist of single characters.
  Outside $S$, $T'$ has $n-h$ occurrences of $\$^2$, to which one can align $\$^2$ in the prefix and the suffix $\$^{2n-2}$ of $P'$.
  This gives $n-h$ segments.
  Each of the remaining $(4n-4)-2(n-h) = 2(n+h-2)$ occurrences of $\$$ in $P'$ is aligned with those in the prefix $(\$0)^{2n-2}$ and the suffix $(0\$)^{2n-2}$ of $T'$.
  In total, we have $P' \in \SegSub{f}(T')$ for $f=m+(n-h)+2(n+h-2) = 3n + m +h - 4$.

  Suppose $P' \in \SegSub{f}(T')$.
  Since $P$ is preceded by $2n$ occurrences of $\$$ in $P'$, $P[1]$ cannot be aligned to any occurrence of $0$ in the prefix $(\$0)^{2n-2}$ of $T'$, which has only $2n-2$ occurrences of $\$$.
  Similarly $P[m]$ cannot be aligned to any $0$ in the suffix $(0\$)^{2n-2}$ of $T'$.
  So, there must be $i$ and $j$ such that $P[1]$ and $P[m]$ are aligned at $T[i]$ and $T[j]$, respectively.
  Obviously, $P \in \SubSeq(T[i..j])$.
  That is, $T$ has a factor of length $j-i+1$ that subsumes $P$ as a subsequence.
  Here, we require $m$ segments to align $P$ with $T[i]\$^2 \cdots \$^2 T[j]$ in $T'$.
  Let us consider how $\$$'s in $P'$ can be aligned with $T'$.
  They cannot be aligned with any of the occurrences of $\$^2$ between $T[i]$ and $T[j]$.
  The occurrence of $\$^2$ immediately before and after $P$ can be put into the segments of $P[1]$ and $P[m]$, respectively.
  This does not require additional segments.
  Now $T'$ has $n-j+i-1$ occurrences of $\$^2$.
  To make the segmentation number as low as possible, we must make each segment as long as possible.
  Thus, we should align as many occurrences of $\$$ in $P'$ as possible with those $n-j+i-1$ occurrences of $\$^2$, which gives $n-j+i-1$ segments, and this leaves $4n-4 - 2(n-j+i-1)$ occurrences of $\$$ in $P'$.
  Those remaining occurrences each will constitute single segments.
  Thus, in total, we have $m + 4n-4 - (n-j+i-1) = m+3n+j-i-3$ segments at minimum.
  Since this number is at most $f = 3n+m+h-4$, the length of $T[i..j]$ is bounded by
  $j-i+1 \le f - (m+3n-4) = h$.
\end{proof}
\begin{example}\label{ex:em2segspm}
  \newcommand{\pt}[1]{\phantom{#1}}
  Consider $T= 0101$ of length $n=4$ and $P=00$ of length $m=2$,
  where $P$ is a subsequence of the factor $T[1..3]$ of length $h=3$ of $T$.
  The proof of Theorem~\ref{th:em2segspm} constructs
  \begin{align*}
    T' &= (\$ 0)^6 \$ \$ 0 \$ \$  1 \$ \$ 0 \$ \$ 1 \$ \$ (0 \$)^6\,,
    \\
    P' & = \$^{8} 0 0 \$^8 \,.
  \end{align*}
  which can be aligned as
  \begin{align*}
    T' &=  \$0\$0\$0\$0\$0\$0 \$ \$ 0 \$ \$  1 \$ \$ 0 \$ \$ 1 \$ \$ 0\$0\$0\$0\$0\$0\$\,,
    \\
    P' & = \$ \pt{1} \$ \pt{1}\$ \pt{1}\$ \pt{1}\$ \pt{1}\$ \pt{1} \$\$ 0 \pt{11111} 0 \$\$ \pt{1} \$\$ \pt{1} \$ \pt{1} \$ \pt{1} \$ \pt{1} \$\,,
  \end{align*}
  where the segmentation number is $f = 13 = 3n+m+h-4$.
\end{example}
\section{Algorithms for {\SegSPM}} \label{sec:SegSPM_algorithms}

In this section, we give a matching upper bound for {\SegSPM}.
We do so by presenting
an algorithm for the function version of the problem, {\SegSEM},
running in $O(mn)$ time, where $m = |P|$ and $n = |T|$.

\subsection{$O(mn)$-time Algorithm for general $f \ge 1$}

Shapira and Storer considered Problem~\ref{def:SegNum} 
in the context of \emph{generalized edit distance problems} with block deletions,
and proposed an $O(mn^2)$-time algorithm~(Algorithm~2 of~\cite{ShapiraS11}).
However, there is a room for improvements in their algorithm,
e.g., employing a standard memorization technique in function $\mathsf{during\_deletion}$ appears to provide $n$ times speedup.
On the other hand, Problem~\ref{def:SegNum} can be seen as a special case of the global alignment with \emph{affine gap penalty},
where the \emph{gap open penalty} equals $1$, the \emph{gap extension penalty} equals $0$, and any gap in $P$ is prohibited.
Our algorithm below is inspired by alignment algorithms with affine gap penalties~\cite{SMITH1981195,GOTOH1982705,Gusfield,Crochemore_Hancart_Lecroq_2007,GiancarloH08}.

\begin{theorem}\label{lem:minseg}
  The problem {\SegSEM} can be solved in $O(|T||P|)$ time.
\end{theorem}
\begin{proof}
  By definition of $\SubSeqf(T)$, if $P \in \SubSeqf(T)$ holds,
  then $P$ is obtained by deleting
  (1) a (possibly empty) prefix of $T$,
  (2) a (possibly empty) suffix of $T$, and
  (3) at most $f-1$ non-empty factors of $T$ that are neither prefix nor suffix
  from $T$.
  The smallest such number $f$ can be computed as $f = d + 1$,
  where $d$ is the minimum number of \textit{edit operations} to obtain $P$ from $T$ by
  (i) deleting a prefix $y$ that changes $yx$ into $x$ with cost $0$,
  (ii) deleting a suffix $y$ that changes $xy$ into $x$ with cost $0$, and
  (iii) deleting a string $y$ that changes $xyz$ into $xz$ with cost $1$,
  where $x, y, z \in \Sigma^{+}$.
  Thus, the standard dynamic programming algorithm~\cite{WagnerFischer1974} can be applied,
  by designing the cost function properly as follows.
  Both the \textit{substitution} and \textit{insertion} operations cost $\infty$ to be prohibited.
  The \textit{block deletion} of $y$ costs $0$ if $y$ is either prefix or suffix, and costs $1$ otherwise.
  Precisely, we can compute the smallest cost $d$ as follows.
  We utilize two tables $D$ and $E$ of size $n \times m$, where $n=|T|$ and $m=|P|$.
  The value $D[i,j]$ indicates the edit distance between $T[1..i]$ and $P[1..j]$
  such that the last operation is a character deletion of $T$, and
  the value $E[i,j]$ gives the edit distance between $T[1..i]$ and $P[1..j]$,
  where $D$ and $E$ both assume that the deletion of a suffix of $T$ costs $1$ instead of $0$.
  These two tables are linked by the following recurrence relations.
  \begin{align*}
    D[i, 0] &= E[i, 0] = 0  \mbox{ \quad for } 0 \leq i \leq n, \\
    D[0, j] &= E[0, j] = \infty  \mbox{ \quad for } 1 \leq j \leq m, \\
    D[i, j] &= \min\{ D[i-1, j], \ E[i-1, j] + 1\},  \mbox{ and}\\  
    E[i, j] &= 
    \begin{cases}
      \min\{ E[i-1, j-1], \ D[i, j] \} & (T[i] = P[j]) \\
      D[i, j]                           & (T[i] \neq P[j]) 
    \end{cases} \\
            & \mbox{ for $1 \leq i \leq n$ and $0 \leq j \leq m$.}
  \end{align*}
  Note that, in the third equation, the ``$+1$'' term indicates the cost to begin a block deletion.
  The smallest cost $d$ is given by $\displaystyle \min_{1 \leq i \leq n}\{ E[i, m] \}$, because we allow to delete a suffix of $T$ with cost $0$.
  If $d = \infty$, it means that $P \notin \SubSeq(T)$.

  The total running time is $O(|T||P|)$.
\end{proof}

\subsection{$O(m+n)$-time Algorithm for $f \le 2$} \label{sec:KMP}

We show a faster algorithm for a special case of {\SegSPM} where the segmentation number $f$ is at most two.
If $f=1$, the problem is nothing more than the standard exact pattern matching, so that it can be solved in $O(|T|+|P|)$ time with $O(1)$ space~\cite{GalilS83,CrochemoreP91}.
Here, we focus on the case where $f=2$.
\begin{theorem}\label{th:segSPM f=2}
  The problem {\SegSPM} with input $f=2$ can be decided in $O(|T|+|P|)$ time with $O(|P|)$ space.
\end{theorem}

For a text $T$ of length $n$ and a pattern $P$ of length $m$, we define three arrays of size $n$ as follows:
\begin{eqnarray*}
  \lpf[i] & = & \max\{ l \mid P[1..l] = T[i-l + 1..i] \}, \\
  \lsf[i] & = & \max\{ l \mid P[m-l+1..m] = T[i..i+l-1] \},\\
  \llpf[i] & = & \max\{ \lpf[j] \mid j \leq i \}, \mbox{\qquad\qquad\qquad for each } 1 \leq i \leq n.
\end{eqnarray*}
The value $\lpf[i]$ (resp. $\lsf[i]$) represents the length of the longest prefix (resp. suffix) of $P$, whose occurrence ends (resp. begins) at position $i$ in $T$. Table~\ref{table:lpf table} shows an example.

\newcolumntype{Y}{>{\centering\arraybackslash}p{0.018\textwidth}}
\begin{table}[tbh]
  \caption{The three arrays $\lpf$, $\lsf$, and $\llpf$ for $T = \texttt{baacababbabcaacaabcba}$ and $P=\texttt{abbabaca}$. In $\llpf$, the \textbf{bold type} emphasizes the values that are greater than their left neighbors.}
  \label{table:lpf table}
  \centering
  \setlength{\arraycolsep}{2pt}
  \begin{tabular}{|c||Y|Y|Y|Y|Y|Y|Y|Y|Y|Y|Y|Y|Y|Y|Y|Y|Y|Y|Y|Y|Y|Y|Y|Y|Y|Y}
    \hline
    $i$ & 1 & 2 & 3 & 4 & 5 & 6 & 7 & 8 & 9 & 10 & 11 & 12 & 13 & 14 & 15 & 16 & 17 & 18 & 19 & 20 & 21\\
    \hline
    $T$ & \ttb & \tta & \tta & \ttc & \tta & \ttb & \tta & \ttb & \ttb & \tta & \ttb & \ttc & \tta & \tta & \ttc & \tta & \tta & \ttb & \ttc & \ttb & \tta \\
    \hline
    $\lpf$ & 0 & 1 & 1 & 0 & 1 & 2 & 1 & 2 & 3 & 4 & 5 & 0 & 1 & 1 & 0 & 1 & 1 & 2 & 0 & 0 & 1\\
    \hline
    \,$\llpf$\, & 0 & \textbf{1} & 1 & 1 & 1 & \textbf{2} & 2 & 2 & \textbf{3} & \textbf{4} & \textbf{5} & 5 & 5 & 5 & 5 & 5 & 5 & 5 & 5 & 5 & 5\\
    \hline
    $\lsf$ & 0 & 1 & 3 & 2 & 1 & 0 & 1 & 0 & 0 & 1 & 0 & 2 & 1 & 3 & 2 & 1 & 1 & 0 & 0 & 0 & 1\\
    \hline
  \end{tabular}
\end{table}

We can easily verify the next lemma.
\begin{lemma}\label{lemma:subseq2}
  The following three conditions are equivalent.
  \begin{itemize}
    \item[(1)] $P \in \SubSeq^{\leq 2}(T)$.
    \item[(2)] $\lpf[i] + \lsf[j] \geq |P|$ for some $1 \leq i < j \leq n$.
    \item[(3)] $\llpf[i] + \lsf[i+1] \geq |P|$ for some $1 \leq i < n$.
  \end{itemize}
\end{lemma}

We now show a sketch of the proof of Theorem~\ref{th:segSPM f=2}. 
At first, we remark that 
by using the Knuth-Morris-Pratt (KMP) automaton~\cite{KMP1977} for $P$, we can compute all values of $\lpf$ from left to right in $O(n)$ time with $O(m)$ space.
Symmetrically, $\lsf$ can be computed from right to left, by KMP automaton for the reverse of $P$.
Moreover, $\llpf$ is easily obtained from $\lpf$.

After constructing these two KMP automata, 
our algorithm consists of the following two phases.
The first phase computes both $\lpf$ and $\llpf$ from left to right. In the second phase, as soon as computing $\lsf$ at each position $i$ from right to left, we check the condition (3) in Lemma~\ref{lemma:subseq2}. If the condition is satisfied, report it and terminate immediately. Otherwise, at the end, we conclude that $P \not\in \SubSeq^{\leq 2}(T)$.
The total running time is $O(n+m)$.
However, the space requirement looks like $O(n)$ due to the three arrays. We reduce it to $O(m)$ as follows. In the second phase, the array $\lsf$ actually does not need to be memorize, because each value is used only once to check the condition.
In the first phase, $\lpf$ is just used to compute $\llpf$, so that we only need $\llpf$ for the second phase. Moreover, remark that the values in $\llpf$ are non-decreasing and range from $0$ to $m$. Thus, we memorize only pairs of the value $\llpf[i]$ and the position $i$ at which $\llpf[i-1] < \llpf[i]$ holds. (See Table~\ref{table:lpf table}, where these values are in bold.) It fits in $O(m)$ space.

\section{Fast Algorithm for {\SegLCS}} \label{sec:Nakatsu}

\newcommand{\ntable}{L}

In this section, we present an efficient algorithm for {\SegLCS}
when the length of the solution is sufficiently long.
Our solution works for any non-fixed $f$ given as input unlike the problem discussed in~\cite{ManeaRS2024}.
We show the following theorem in this section:
\begin{theorem} \label{th:lcsf}
  Given strings $T_1$ and $T_2$ with $|T_1| \le |T_2|$ over a linearly sortable alphabet, and integer $f$,
  the problem {\SegLCS} can be solved in $O(f|T_2|(|T_1|-\ell+1))$ time using $O(\ell(|T_1|-\frac{\ell}{f}+1))$ space, where $\ell = \LCSk(T_1, T_2, f)$.
\end{theorem}

The basic idea of our algorithm is to combine the ideas from Banerjee et al.'s algorithm~\cite{BanerjeeGT24} for the {\SegLCS} problem and
the method of Nakatsu et al.~\cite{nakatsu1982longest} for the (standard) LCS problem. 
Banerjee et al.~\cite{BanerjeeGT24} considered
a function\footnote{We remark that the definition of $C$ here is not strictly the same as the corresponding function $\mathsf{CHAIN}$ from~\cite{BanerjeeGT24} as $C$ focuses on prefixes of $T_1$ and $T_2$ while $\mathsf{CHAIN}$ focuses on suffixes of them.} $C$ such that
$C(i,j,h) = \LCSk(T_1[1..i], T_2[1..j], h)$
for $1 \leq i \leq n_1$, $1 \leq j \leq n_2$, and $1 \leq h \leq f$.
They showed that $C$ can be computed by the following recurrence:
\begin{equation*}
  C(i,j,h) = \max\{C(i, j - 1, h), C(i - 1, j, h), Z\},
\end{equation*}
where $Z = x + C(i - x, j - x, h-1)$ and $x=\lcsuf(T_1[1.. i], T_2[1..j])$.
This recurrence provides an $O(fn_1n_2)$-time algorithm since $\lcsuf(T_1[1.. i], T_2[1..j])$ can be computed in $O(1)$ time after $O(n_1+n_2)$-time preprocessing by using a \emph{lowest common ancestor} query~\cite{LCE_article} on the \emph{suffix tree}~\cite{Weiner73,farach1997optimal} of the reversal of $T_1T_2$.
We further speed up the solution by combining Nakatsu et al.'s method.

Our algorithm maintains a three-dimensional table.
Let $\ntable(i,s,h)$ be the length $j$ of the shortest prefix $T_2[1..j]$ of $T_2$ satisfying $\LCSk(T_1[1..i], T_2[1..j], h) = s$. If there are no such prefixes, let $\ntable(i, s, h) = \infty$.
Then, we have
\[
  \LCSk(T_1,T_2,f) = \max\{\,s \mid \ntable(i, s, f) \neq \infty \text{ for some } i\,\}
\,.\]
Our goal is to compute the table $\ntable$ by dynamic programming, but as we will see later, we do not have to completely fill the table to obtain the value $\LCSk(T_1,T_2,f)$.

We have the following inequalities by definition.
\begin{align} 
  \ntable(i, s, h) &\leq \ntable(i-1, s, h)\label{eq:2}\\
  \ntable(i, s, h) &> \ntable(i-1, s-1, h)\label{eq:3}
\end{align}

Also, we obtain the following lemmas.
\begin{lemma} \label{lem:j_isk}
  For any $i, j, s$, and $h$, if $j\geq \ntable(i-x, s-x, h-1)+x$ holds where $x = \lcsuf(T_1[1..i],T_2[1..j])$, then $\ntable(i, s, h) \leq j$ holds. 
\end{lemma}
\begin{proof}
  Since $j-x \ge \ntable(i-x, s-x, h-1)$, $T_1[1..i-x]$ and $T_2[1..j-x]$ have $(h-1)$-SegCS of length $s-x$.
  Thus, $T_1[1..i]$ and $T_2[1..j]$ have $h$-SegCS of length $s$ since $x = \lcsuf(T_1[1..i],T_2[1..j])$.
  Therefore, $\ntable(i, s, h) \leq j$ holds.
\end{proof}

The next lemma states that, intuitively, a right-aligned embedding of a string into $S$ and $T$ can be transformed into another embedding while maximizing the length of the rightmost segment without increasing the segmentation size.
\begin{lemma}\label{lem:LCS_has_LCE}
  Suppose $p = (u_1,\dots,u_h)$ is an $h$-segmentation of a string of length $s = |u_1 \dots u_h|$
  such that $p$ can be embedded to both of $S$ and $T$
  and the last segment $u_h$ is a common suffix of $S$ and $T$.
  Then, there is an $h$-segmentation $p' = (u_1',\dots,u_h')$ of a string of length $s$ that can be embedded to both $S$ and $T$
  such that $|u_h'| = \min \{ \lcsuf(S,T), s \}$.
\end{lemma}
\begin{proof}
  Let $\ell = \lcsuf(S,T)$.
  If $\ell \ge s$, then the $h$-segmentation $(\varepsilon,\dots,\varepsilon, S[|S|-s+1,|S|])$ of $S[|S|-s+1,|S|] = T[|T|-s+1..|T|]$ has the desired property.   
  If $\ell < s$, let $j$ be the integer such that $|u_1 \dots u_{j-1}| < s - \ell \le |u_{1} \dots u_{j}|$.
  Such $j$ always exists. Also, $j < h$ holds since $|u_h| \leq \ell$.
  Let $v$ be the prefix of $u_{j}$ such that $|u_1 \dots u_{j-1} v| = s- \ell$.
  Let $u_h' = S[|S|- \ell +1,|S|]$.
  Then, the sequence of $h$ segments $(u_1 ,\dots, u_{j-1}, v,\varepsilon,\dots,\varepsilon,u_h')$ has the desired property.
\end{proof}

The following lemma is the core of our algorithm.
\begin{lemma} \label{lem:re}
  \[
    \ntable(i,s,h) = \min\{\ntable(i-1, s, h), j_{i, s, h}\}
  \]
  where
  \[
    j_{i,s,h} = \min\{j \mid j \ge \ntable(i - x, s - x, h - 1) + x \text{ for } x = \min\{\lcsuf(T_1[1..i],T_2[1..j]),s\}\}\,.
  \]
\end{lemma}
\begin{proof}
  By Equations~(\ref{eq:2}) and Lemma~\ref{lem:j_isk}, $\ntable(i,s,h) \le \min \{\ntable(i-1, s, h), j_{i, s, h}\}$ holds.
  Suppose $\ntable(i,s,h) < \ntable(i-1, s, h)$.
  Then, the last segment of an $h$-segmentation witnessing $\ntable(i,s,h) = j$ must involve the last letter of $T[1..i]$.
  By Lemma~\ref{lem:LCS_has_LCE}, there is an $h$-segmentation witnessing $\ntable(i,s,h) = j$ whose last segment has length $x=\min\{\lcsuf(T_1[1..i],T_2[1..j]),s\}$ and the other $h-1$ segments are embedded into $T_1[1..i-x]$ and $T_2[1..j-x]$.
  That is, $\ntable(i-x,s-x,h-1) \le j-x$, i.e., 
  \[
    \ntable(i,s,h) = j \ge \ntable(i-x,s-x,h-1)+x \ge j_{i,s,h}
  \,.\]
  Therefore, $\ntable(i,s,h) = j_{i,s,h}$ if $\ntable(i,s,h) < \ntable(i-1, s, h)$.
\end{proof}
Our algorithm computes values in the table $\ntable$ based on Lemma~\ref{lem:re}.

Let $\ntable_h$ be the $h$-th table which stores $\ntable(i,s,h)$ for $1 \leq i,s \leq |T_1|$. 
Here, $\LCSk(T_1,T_2,f)$ is the largest $s$ such that $\ntable(i, s, f) \neq \infty$ for $ 0 \leq i \leq |T_1|$, 
i.e., the $s$-th row is the lowest row which has an integer value in the $\ntable_f$. 
We show an example of three-dimensional table $\ntable$ for $T_1 = \mtt{abcabbac}$, $T_2 = \mtt{bcbcbbca}$ and $f = 3$ in 
Table~\ref{tab:table_1}.

\begin{table}[tbp]
  \centering
  \caption{An example of three-dimensional table $\ntable = (\ntable_1, \ntable_2, \ntable_3)$ for strings $T_1 = \mtt{abcabbac}$, $T_2 = \mtt{bcbcbbca}$ and $f = 3$. From left to right, the tables are $\ntable_1$, $\ntable_2$, and $\ntable_3$.}
  \small
  \renewcommand{\arraycolsep}{2pt}
  \begin{tabular}{@{}l@{}l@{}l}
    \begin{minipage}{0.33\textwidth}
      \centering
      \[
        \begin{array}{c|cccccccc}
          s \backslash i & 1 & 2 & 3 & 4 & 5 & 6 & 7 & 8 \\
          \hline
          1 & 8 & 1 & 1 & 1 & 1 & 1 & 1 & 1 \\
          2 & \infty & \infty & 2 & 2 & 2 & 2 & 2 & 2 \\
          3 & \infty & \infty & \infty & 8 & 8 & 8 & 8 & 8 \\
          4 & \infty & \infty & \infty & \infty & \infty & \infty & \infty & \infty \\
          5 & \infty & \infty & \infty & \infty & \infty & \infty & \infty & \infty \\
          6 & \infty & \infty & \infty & \infty & \infty & \infty & \infty & \infty \\
          7 & \infty & \infty & \infty & \infty & \infty & \infty & \infty & \infty \\
          8 & \infty & \infty & \infty & \infty & \infty & \infty & \infty & \infty \\
        \end{array}
      \]
    \end{minipage} &
    \begin{minipage}{0.33\textwidth}
      \centering
      \[
        \begin{array}{c|cccccccc}
          s \backslash i & 1 & 2 & 3 & 4 & 5 & 6 & 7 & 8 \\
          \hline
          1 & 8 & 1 & 1 & 1 & 1 & 1 & 1 & 1 \\
          2 & \infty & \infty & 2 & 2 & 2 & 2 & 2 & 2 \\
          3 & \infty & \infty & \infty & 8 & 3 & 3 & 3 & 3 \\
          4 & \infty & \infty & \infty & \infty & \infty & 6 & 6 & 6\\
          5 & \infty & \infty & \infty & \infty & \infty & \infty & \infty & \infty \\
          6 & \infty & \infty & \infty & \infty & \infty & \infty & \infty & \infty \\
          7 & \infty & \infty & \infty & \infty & \infty & \infty & \infty & \infty \\
          8 & \infty & \infty & \infty & \infty & \infty & \infty & \infty & \infty \\
        \end{array}
      \]
    \end{minipage} &
    \begin{minipage}{0.33\textwidth}
      \centering
      \[
        \begin{array}{c|cccccccc}
          s \backslash i & 1 & 2 & 3 & 4 & 5 & 6 & 7 & 8 \\
          \hline
          1 & 8 & 1 & 1 & 1 & 1 & 1 & 1 & 1 \\
          2 & \infty & \infty & 2 & 2 & 2 & 2 & 2 & 2 \\
          3 & \infty & \infty & \infty & 8 & 3 & 3 & 3 & 3 \\
          4 & \infty & \infty & \infty & \infty & \infty & 5 & 5 & 4 \\
          5 & \infty & \infty & \infty & \infty & \infty & \infty & 8 & 7 \\
          6 & \infty & \infty & \infty & \infty & \infty & \infty & \infty & \infty \\
          7 & \infty & \infty & \infty & \infty & \infty & \infty & \infty & \infty \\
          8 & \infty & \infty & \infty & \infty & \infty & \infty & \infty & \infty \\
        \end{array}
      \]
    \end{minipage}
  \end{tabular}
  \label{tab:table_1}
\end{table}

Next, we introduce the main part of our algorithm.
In our algorithm, we compute the values for each diagonal line from upper left to lower right in left-to-right order. 
Then, we do not compute all values in the tables, since there are unrelated values for obtaining the length of a longest $f$-SegCS. 
While computing a diagonal line from upper-left to lower-right,
we can stop the computation when we meet $\infty$ since the values of the remaining cells on the line are guaranteed to be $\infty$ by Equation~(\ref{eq:3}). 
This is reflected to lines~\ref{code:lines_break_2} and~\ref{code:return} of Algorithm~\ref{alg:sub} we will see later.
Also, while processing diagonal lines from left to right in table $\ntable_h$,
we can stop the iteration
once we reach a cell in the rightmost column of $\ntable_h$ in the computation of some $p$-th diagonal.
This is because, for any $q > p$, the lowest endpoint of $q$-th diagonal must be higher than that of $p$-th diagonal, and thus, computing such diagonal is meaningless for our purpose.
This is reflected to line~\ref{code:table_break_1} of Algorithm~\ref{alg:1} we will see later.

From the above, we propose an algorithm shown in Algorithms~\ref{alg:1} and~\ref{alg:sub}.
For convenience, we assume that the values of uninitialized $\ntable(\cdot, \cdot, \cdot)$ are $\infty$.
The main procedure is shown in Algorithm~\ref{alg:1}.
We compute tables $\ntable_h$ for incremental $h = 1, \ldots , f$.
In the while-loop, we compute the $\mathit{diag}$-th diagonal lines of table $\ntable_h$ for $\mathit{diag} = 0, 1, 2, \ldots$, i.e., from left to right.
The subroutine to fill a diagonal line is denoted by $\mathit{FillDiagonally}$, which is shown in Algorithm~\ref{alg:sub}.
Here, we introduce an array $\LowestRowNum$ of size $f$ such that,
while computing each table $\ntable_h$ in the while-loop, $\LowestRowNum[h]$ keeps the largest vertical index $s$ satisfying that the $s$-th row of table $\ntable_h$ has a finite value.
Thus, at the end of the computation of the last table $\ntable_f$~(at line~\ref{code:last}), $\LCSk(T_1, T_2, f) = \LowestRowNum[f]$ holds.
By the definition of $\LowestRowNum[h]$, the condition of the while-statement is valid
since for every $e \ge n_1 - \LowestRowNum[h]$, the $e$-th diagonal cannot reach the $\LowestRowNum[h]$-th row as discussed above.

In subroutine $\mathit{FillDiagonally}$~(Algorithm~\ref{alg:sub}),
we fill the $\mathit{diag}$-th diagonal line from upper-left to lower-right,
where $s$ is a vertical index and $i$ is a horizontal index of the table $\ntable_h$.
Each cell in the diagonal line is computed in the while-loop.
Lines~\ref{code:computex}--\ref{code:endif} are due to Lemma~\ref{lem:re}.
Also, if $\ntable(i,s,h)$ is not updated in the while-loop, its value is $\infty$, and thus, we update $\LowestRowNum[h] \leftarrow s-1$.
Further, as discussed above, we do not need to compute below $s$ on the diagonal line anymore since they are all $\infty$.
Then we terminate the procedure.

We give an example of the sparse table computed by Algorithms~\ref{alg:1} and~\ref{alg:sub} for input strings $T_1 = \mtt{abcabbac}$, $T_2 = \mtt{bcbcbbca}$ and $f = 3$ in 
Table~\ref{tab:table_2}.
For example, in the third diagonal line of $\ntable_2$, let us consider computing $\ntable(6, 4, 2)$ and assume that $\ntable(5, 3, 2)$ has already been computed. 
We can start searching the index of $T_2$ satisfying the condition of recurrence from $\ntable(5, 3, 2) + 1 = 4$. 
Then, the smallest index $6$ of $T_2$ satisfying $\lcsuf(T_1[1..6], T_2[1..6]) = 2$ and $\ntable(6-2, 4-2, 1) = 2 < 6$ is the answer of $\ntable(6, 4, 2)$.

\begin{table}[tbp]
  \centering
  \caption{An example of the sparse table computed by Algorithm~\ref{alg:1} for input strings $T_1 = \mtt{abcabbac}$, $T_2 = \mtt{bcbcbbca}$ and $f = 3$. From left to right, the tables are $\ntable_1$, $\ntable_2$, and $\ntable_3$.}
  \small
  \renewcommand{\arraycolsep}{2pt}
  \begin{tabular}{@{}l@{}l@{}l}
    \begin{minipage}{0.3\textwidth}
      \centering
      \[
        \begin{array}{c|CCCCCCCC}
          s \backslash i & 1 & 2 & 3 & 4 & 5 & 6 & 7 & 8 \\
          \hline
          1 & 8 & 1 & 1 & 1 & 1 &  &  &  \\
          2 & & \infty & 2 & 2 & 2 & 2 &  &  \\
          3 & & & & 8 & 8 & 8 & 8 &  \\
          4 & & & & & \infty & \infty & \infty & \infty \\
          5 & & & & & & & & \\
          6 & & & & & & & & \\
          7 & & & & & & & & \\
          8 & & & & & & & & \\
        \end{array}
      \]
    \end{minipage} &
    \begin{minipage}{0.3\textwidth}
      \centering
      \[
        \begin{array}{c|CCCCCCCC}
          s \backslash i & 1 & 2 & 3 & 4 & 5 & 6 & 7 & 8 \\
          \hline
          1 & 8 & 1 & 1 & 1 &  &  &  &  \\
          2 & & \infty & 2 & 2 & 2 &  &  &  \\
          3 & & & & 8 & 3 & 3 &  &  \\
          4 & & & & & \infty & 6 & 6 & \\
          5 & & & & & & & \infty & \infty \\
          6 & & & & & & & & \\
          7 & & & & & & & & \\
          8 & & & & & & & & \\
        \end{array}
      \]
    \end{minipage} &
    \begin{minipage}{0.3\textwidth}
      \centering
      \[
        \begin{array}{c|CCCCCCCC}
          s \backslash i & 1 & 2 & 3 & 4 & 5 & 6 & 7 & 8 \\
          \hline
          1 & 8 & 1 & 1 &  &  &  &  &  \\
          2 & & \infty & 2 & 2 &  &  &  &  \\
          3 & & & & 8 & 3 &  &  &  \\
          4 & & & & & \infty & 5 &  &  \\
          5 & & & & & & & 8 &  \\
          6 & & & & & & & & \infty \\
          7 & & & & & & & & \\
          8 & & & & & & & & \\
        \end{array}
      \]
    \end{minipage}
  \end{tabular}
  \label{tab:table_2}
\end{table}

\begin{algorithm}[tbp]
  \caption{$O(fn_1(n_2-\ell+1))$-time algorithm for {\SegLCS}}
  \label{alg:1}
  \begin{algorithmic}[1]    
    \Require Strings $T_1$ of length $n_1$ and $T_2$ of length $n_2$, and a positive integer $f$
    \Ensure $\LCSk(T_1, T_2, f)$
    \For{$h \leftarrow 1$ to $f$} \label{code:for_k}
    \State $\LowestRowNum[h] \leftarrow 0$
    \State $\mathit{diag} = 0$
    \While {$\mathit{diag} < n_1 - \LowestRowNum[h]$} \label{code:table_break_1}
    \State $\ntable(\mathit{diag},0,h) \leftarrow 0$
    \State $\mathit{FillDiagonally}(h, \mathit{diag}, \LowestRowNum)$ \Comment{May update $\LowestRowNum[h]$.} \label{code:callsub}
    \State $\mathit{diag} \leftarrow \mathit{diag} + 1$ 
    \EndWhile 
    \EndFor \Comment{$\LowestRowNum[h] = \LCSk(T_1, T_2, h)$ holds at this line.}
    \State \Return $\LowestRowNum[f]$ \label{code:last}
  \end{algorithmic}
\end{algorithm}

\begin{algorithm}[tbp]
  \caption{Subroutine $\mathit{FillDiagonally}(h, \mathit{diag}, \LowestRowNum)$}
  \label{alg:sub}
  \begin{algorithmic}[1]
    \State $j \leftarrow 1$
    \For{$s \leftarrow 1$ to $n_1 - \mathit{diag}$} \label{code:for_s_1}
    \State $i \leftarrow s + \mathit{diag}$
    \State $\ntable(i, s, h) \leftarrow \infty$ \Comment{Initialize.}
    \While{$j \leq n_2$} 
    \State{$x \leftarrow \lcsuf(T_1[1..i], T_2[1..j])$}\label{code:computex}
    \If{$j = \ntable(i-1, s, h)$ or $x > 0$ and $j \geq x + \ntable(i - x, s-x, h-1)$}
    \State $\ntable(i, s, h) \leftarrow j$ \Comment{By Lemma~\ref{lem:re}.}
    \State \textbf{break}
    \EndIf\label{code:endif}
    \State $j \leftarrow j + 1$ 
    \EndWhile 
    \If{$\ntable(i, s, h) = \infty$} \label{code:lines_break_2}
    \State $\LowestRowNum[h] \leftarrow s-1$
    \State \Return \label{code:return}
    \EndIf
    \State $j \leftarrow j+1$
    \EndFor \label{code:for_s_2}
    \State \Return
  \end{algorithmic}
\end{algorithm}

Finally, we discuss the complexity of our algorithm. 
For each table $\ntable_h$, our algorithm computes
at most $n_1 - \ell_h$ diagonal lines
where $\ell_h$ is the largest $s$ such that $\ntable(i, s, h) < \infty$ for some $i$, which is equivalent to the final value of $\LowestRowNum[h]$.
Also, for each diagonal line in $\ntable_h$,
at most $\ell_h + 1$ cells are accessed while processing the line.
Further, such cells can be computed in $O(n_1 - \mathit{diag} + n_2) = O(n_2)$ time by $\mathit{FillDiagonally}$
since the value of the positive integer $j$ is incremented by $1$ at the end of the inner while-loop and $j$ is upper-bounded by $n_2$.
Hence the computation time to fill the table $\ntable_h$ is $O(n_2(n_1 - \ell_h+1))$ and the size of $\ntable_h$ is $O(\ell_h(n_1 - \ell_h+1))$.
Thus, the total time complexity can be written as $O(\sum_{h=1}^{f} n_2(n_1 - \ell_h +1))$.
From the definition of $\SegLCS$,
$\ell_1 \geq \ell/f$ holds since $\ell = \LCSk(T_1, T_2, f)$ and $\ell_1$ is the length of the longest common factor of $T_1$ and $T_2$.
Similarly, $\ell_h \geq h\ell/f$ holds for $1 \leq h \leq f$.
Thus, the total time complexity is $O(fn_2(n_1-\ell+1))$ since
$\sum_{h=1}^{f} n_2(n_1 - \ell_h + 1) \le \sum_{h=1}^{f} n_2(n_1 - h\ell/f + 1) = fn_2(n_1 - \ell/2 + 1) - n_2\ell/2$.

Finally, the total size of tables we actually use is $O(\max_h\{\ell_h(n_1 - \ell_h+1)\})$ since our algorithm can compute each table $\ntable_h$ only using the values in at most two tables $\ntable_h$ and $\ntable_{h-1}$.
Again, since $\ell_h \geq h\ell/f \geq \ell/f$ holds for each $1 \le h \le f$,
the total space complexity is $O(\ell(n_1 - \ell/f + 1))$.
We note that, although we depict table $\ntable_h$ as a two-dimensional table in our examples,
we can easily implement each table as one-dimensional array of size $O(\ell_h(n_1 - \ell_h+1))$ representing the diagonal lines.

To summarize, we obtain Theorem~\ref{th:lcsf}.

\section{Conclusions and discussions} \label{sec:conclusions}

In this paper, we have studied two basic problems concerning $f$-segmental subsequences:
the pattern matching problem and the LCS problem.
Concerning the former, we have presented 
a quadratic-time algorithm for general $f \ge 1$ and a linear-time algorithm for $f \le 2$. 
We also argued that the quadratic time complexity is unavoidable under the {strong exponential-time hypothesis}.
The latter problem has already been studied in the literature 
The $f$-segmental LCS problem and its variants have already been studied in the literature~\cite{LiZJFC22,LiJCFZ24,BanerjeeGT24}.
Our algorithm runs faster than theirs when the solution is long.

Studying mathematical and computational properties of subsequences formed by a limited number of segments is relatively a new topic.
Below, we raise a few open questions related to the properties of $f$-segmental subsequences among many possible research directions.
\begin{itemize}
  \item
    Is it possible to decide whether $P \in \SubSeqf(T)$ more efficiently than $O(mn)$ time for general $f$?
    Our proof of the conditional lower bound uses $f \in \Theta(m)$,
    while Banerjee et al.~\cite{BanerjeeGT24} showed that it can be determined in $\tilde{O}((n_1n_2)^{1-(1/3)^{f-2}})$ time
    for a \emph{constant} $f \geq 3$.
    For example, would an $O(fn)$-time algorithm exist?
  \item 
    Does the conditional lower bound on the time complexity of {\SegSPM} (Theorem~\ref{th:em2segspm}) hold for binary alphabets?
    We are somewhat optimistic on this question.
    Using a similar technique by Bille et al.~\cite[Theorem~1]{Bille2022}, one may be able to reduce the alphabet size, where we might need to use a direct reduction from the orthogonal vector problem, from which Bille et al.\ have shown the conditional lower bound on episode matching.
  \item
    Can we enumerate ``minimal occurrences'' of $P$ in $T$ as subsequences with at most $f$ segments as in the episode matching, for non-constant $f$?
    Banerjee et al.~\cite{BanerjeeGT24} showed it is possible to find one of such occurrences in
    $\tilde{O}((n_1n_2)^{1-(1/3)^{f-2}})$ time for a \emph{constant} $f \geq 3$.
  \item
    Can we bound the size of \emph{subsequence DFAs} for {\SegSPM}?
    Can we efficiently construct such DFAs?
    One can see that a DFA accepting exactly $\SegSub{f}(T)$ requires $\Theta(f|T|)$ states when $T = (\mtt{ab})^n$ and $f \le n/2$, but we do not know if there are any texts for which DFAs require more states than $\Theta(f|T|)$. 
  \item Is it possible to efficiently construct an indexing structure for $T$ on which one can efficiently decide whether $P$ is an $f$-segmental subsequence of $T$?
    The above observation on the DFA size bound does not necessarily refute a possibility of a more elaborated indexing structure of size $O(|T|)$. 
  \item One may consider another generalization of the LCS problem for $f$-segmental subsequences, where we are interested in the length of a longest common $f$-segmental subsequences, i.e., the length of longest elements of $\SegSub{f}(T_1) \cap \SegSub{f}(T_2)$.
    This is a different problem from the one the literature and this paper have discussed.
    For example, $\mtt{abc}$ is a 2-segmental subsequence of both $T_1 = \mtt{abac}$ and $T_2 = \mtt{acbc}$, while they have different witness segmentations $(\mtt{ab},\mtt{c})$ and  $(\mtt{a},\mtt{bc})$.
    A naive dynamic programming algorithm is possible for this problem (Appendix~\ref{sec:seglcs}), but is it possible to design a more efficient algorithm that runs fast when the solution is big enough like our algorithm for \textsf{SegLCS}?
\end{itemize}

\clearpage
\appendix

\section{Algorithm for {\GenSegLCS}}\label{sec:seglcs}

For a non-empty finite set $X$ of strings, define $\maxlen(X)$ to be the length of a longest element of $X$.
If $X$ is empty, let $\maxlen(X) = -\infty$.

In this appendix, we consider the following alternative generalization of the LCS problem.
\begin{problem}[\GenSegLCS]\label{def:GenSegLCS}
  Given two strings $T_1$ and $T_2$ and two positive integers $f_1$ and $f_2$, 
  compute $\maxlen(\SegSub{f_1}(T_1) \cap \SegSub{f_2}(T_2))$.
\end{problem}
This problem {\GenSegLCS} is different from \SegLCS{}, even when $f_1 = f_2$.
Consider $T_1 = \mtt{abcxdexf}$ and $T_2 = \mtt{ab y cdef}$.
The longest 2-SegLCSs of $T_1$ and $T_2$ are $\mtt{abde}$, which is witnessed by the 2-segmentation $(\mtt{ab},\mtt{de})$.
On the other hand, in {\GenSegLCS}, embeddings to $T_1$ and $T_2$ are independent.
The string $\mtt{abcde}$ has two 2-segmentations $(\mtt{abc},\mtt{d})$ and and $(\mtt{ab},\mtt{cd})$, which are embedded to $T_1$ and $T_2$, respectively.
One may consider a more general case where $f_1 \neq f_2$.
For example, $\mtt{abcdef} \in \SegSub{3}(T_1) \cap \SegSub{2}(T_2)$ is considered in {\GenSegLCS}.

This section shows the following theorem.
\begin{theorem}
  The problem {\GenSegLCS} can be solved in $O(g_1g_2 n_1 n_2)$ time for input $T_1,T_2,f_1,f_2$, where $n_1 = |T_1|$, $n_2 = |T_2|$, $g_1 = \min\{f_1,\max\{\lrceil{|T_1|/2}-f_1,1\}\}$, and $g_2 = \min\{f_2,\max\{\lrceil{|T_2|/2}-f_2,1\}\}$.
\end{theorem}
Particularly when $f_1 = \lrceil{n_1/2}$ and $f_2 = \lrceil{n_2/2}$, the {\GenSegLCS} problem is no different than the standard LCS problem and the time complexity of our algorithm becomes $O(n_1 n_2)$.

The basic idea for small segmentation numbers is quite straightforward.
We maintain four\linebreak
four-dimensional dynamic programming tables, among which $L_\mrm{BB}$ records the values\linebreak
$\maxlen(\SegSub{h_1}(T_1[1..i_1]) \cap \SegSub{h_2}(T_2[1..i_2]) )$ for $0 \le i_a \le n_a$ and $0 \le h_a \le f_a$ with $a=1,2$.
By definition, this gives the answer when $(i_1,i_2,h_1,h_2)=(n_1,n_2,f_1,f_2)$.
When extending a current LCS candidate, we should pay attention on whether we start a new segment or extend the last segment.
Define
\[
  \SegSuf{f}(T) = \{\, u_1\cdots u_f \mid T = v_0 u_1 v_1 \dots v_{f-1} u_f \text{ for some } v_0,\ldots, v_{f-1} \in \Sigma^* \,\}
\,,\]
which is a subset of $\SegSub{f}(T)$, with the restriction that the last segment must be a suffix of $T$.
Obviously, $\SegSub{f-1}(T) \subseteq \SegSuf{f}(T) \subseteq \SegSub{f}(T)$.
By maintaining the following four tables, one can solve {\GenSegLCS}.
\begin{align*}
  L_\mrm{BB}[i_1,i_2,h_1,h_2] &= \maxlen( \SegSub{h_1}(T_1[1..i_1]) \cap \SegSub{h_2}(T_2[1..i_2]) ),
  \\
  L_\mrm{BF}[i_1,i_2,h_1,h_2] &= \maxlen( \SegSub{h_1}(T_1[1..i_1]) \cap \SegSuf{h_2}(T_2[1..i_2]) ),
  \\
  L_\mrm{FB}[i_1,i_2,h_1,h_2] &= \maxlen( \SegSuf{h_1}(T_1[1..i_1]) \cap \SegSub{h_2}(T_2[1..i_2]) ),
  \\
  L_\mrm{FF}[i_1,i_2,h_1,h_2] &= \maxlen( \SegSuf{h_1}(T_1[1..i_1]) \cap \SegSuf{h_2}(T_2[1..i_2]) ).
\end{align*}
If any of $i_1,i_2,h_1,h_2$ is zero, then $L_{*}[i_1,i_2,h_1,h_2] = 0$ holds for all \linebreak
$* \in \{\mrm{BB},\mrm{BF},\mrm{FB},\mrm{FF}\}$.
One can easily compute $L_*[i_1,i_2,h_1,h_2]$ inductively for $i_1,i_2,h_1,h_2 > 0$.

Now, we present an improvement on the naive algorithm particularly for the case where the segmentation number $f_a$ is close to $n_a/2$ (but not larger than $\lrceil{n_a/2}$).
Consider a factorization $(v_0, u_1, v_1, \dots, u_h, v_h)$ of $T = v_0 u_1 v_1 \dotsm u_h v_h$ that witnesses $u_1 \dotsm u_h \in \SegSub{f}(T)$ with $h \le f$.
To make the segmentation number $h$ small, some of the factors $u_i$ and $v_i$ must be long enough.
More formally, $u_1 \dotsm u_h \in \SegSub{f}(T)$ if and only if $\mcal{S}(v_0,u_1,\dots,u_h,v_h) \ge |T|-2f$ for the \emph{score} $\mcal{S}$ of the factorization defined by
\begin{equation*}
  \mcal{S}(w_0,\dots,w_m) = |w_0| + \sum_{j=1}^m (|w_i|-1)
  \,.\label{eq:lcs}
\end{equation*}
When extracting a subsequence scanning $T$ from left to right, instead of keeping track of the segmentation number $h$, we memorize the score of a factorization that yields the subsequence.
Once the score of a factorization of a prefix of $T$ reaches $|T|-2f$, any subsequence of the suffix may follow, so we do not have to compute the exact score any further.
When the segmentation constraint $f$ is close to $|T|/2$, the upper bound $|T|-2f$ will be close to zero.

We remark that when the last segment is a suffix of $T$, i.e., $T = v_0 u_1 \dotsm v_{h-1} u_h$, we have $\mcal{S}(v_0,u_1,\dots,v_{h-1},u_h) = |T|-2h+1$.
This appears that we need a higher score than the previous case, where the factorization ends with $v_h$.
But one also notices that, in this case, the score and the text length always have opposite parities.
Therefore, if $\mcal{S}(v_0,u_1,\dots,v_{h-1},u_h) \ge |T|-2f$, then $\mcal{S}(v_0,u_1,\dots,v_{h-1},u_h) \ge |T|-2f+1$.
Hence, we can use the same score threshold $|T|-2f$ for a factorization ending with the last segment $u_h$ to judge whether the yielded subsequence is $f$-segmental.

Define
\begin{align*}
  \Segb{\ell}(T) = \{\,& u_1 \dotsm u_h  \mid  T = v_0 u_1 \dotsm v_{h-1} u_h v_h \text{ for some } v_0 \in \Sigma^* \text{ and } \\ & u_1,v_1,\dots,u_h,v_h \in \Sigma^+
  \text{ with }  \mcal{S}(v_0,u_1,\dots,u_h,v_h) \ge \ell \,\}
  \,,\\
    \Segf{\ell}(T) = \{\, & u_1 \dotsm u_h  \mid T = v_0 u_1 \dotsm v_{h-1} u_h \text{ for some } v_0 \in \Sigma^* \text{ and } \\ & u_1,v_1,\dots,u_h \in \Sigma^+
    \text{ with } \mcal{S}(v_0,u_1,\dots,v_{h-1},u_h) \ge \ell \,\}
    \,.
    \end{align*}
    We remark that we disallow $u_i$ and $v_i$ with $i>0$ to be empty in the definition above, differently from $\SegSub{h}$ and $\SegSuf{h}$.
    In this way, the scores are always non-negative.

    Depending on whether $f_a$ is small or large, we use different types of dynamic programming tables.
    For $X_1,X_2 \in \{\, \mrm{B,F,\ol{B},\ol{F}}\,\}$, we consider the tables $L_{X_1X_2}$,
    which shall satisfy
    \begin{equation}\label{eq:indseg_recurrence}
      L_{X_1X_2}[i_1,i_2,p_1,p_2] = \maxlen \big( S_1(T_1[1..i_1]) \cap S_2(T_2[1..i_2]) \big)
    \end{equation}
    where
    \[
      S_a = \begin{cases}
        \SegSub{p_a}  &\text{if $X_a = \mrm{B}$,}
        \\  \SegSuf{p_a}  &\text{if $X_a = \mrm{F}$,}
        \\  \Segb{p_a}  &\text{if $X_a = \mrm{\ol{B}}$,}
        \\  \Segf{p_a}  &\text{if $X_a = \mrm{\ol{F}}$}
      \end{cases}
    \]
    for $a = 1,2$.
    If $f_a$ is small, say $f_a \le n_a / 4$, we are interested only in $X_a \in \{\mrm{B,F}\}$ with $0 \le i_a \le n_a$ and $0 \le p_a \le f_a$;
    if $f_a$ is large, say $f_a > n_a / 4$, we are interested only in $X_a \in \{\mrm{\ol{B},\ol{F}}\}$ with $0 \le i_a \le n_a$ and $0 \le p_a \le \max\{0,n_a - 2f_a\}$.
    For example, if $f_1$ is small and $f_2$ is large, then we construct the four tables $L_{\mrm{B\ol{B}}}$, $L_{\mrm{B\ol{F}}}$, $L_{\mrm{F\ol{B}}}$, $L_{\mrm{F\ol{F}}}$.
    The answer will be found as the maximum of the $(n_1,n_2,f_1,\max\{0,n_2-2f_2\})$-entries of those tables.
    Hereafter, let $g_a = f_a$ if $f_a$ is small, and $g_a = \max\{0,n_a-2f_a\}$ if $f_a$ is large.
    It remains to explain how to initialize and update those tables.

    \paragraph*{Initialization.}
    For $i_1=0$ or $i_2=0$, let $L_{X_1X_2}[i_1,i_2,p_1,p_2] = \min\{q_1,q_2\}$ where
    \[
      q_a = \begin{cases}
        0 &\text{if $X_a = \mrm{B}$ or $X_a =\mrm{F} \wedge p_a > 0$ or $X_a = \mrm{\ol{B}} \wedge p_a \le i_a$,}
        \\  -\infty &\text{otherwise}
      \end{cases}
    \]
    for $a=1,2$.
    This is justified by the fact that for any $p \ge 0$,
    \begin{gather*}
      \lambda \in \SegSub{p}(\lambda) \cap \SegSuf{1+p}(\lambda) \cap \Segb{p}(T[1..p]) \,,
      \\
      \lambda \notin \SegSuf{0}(\lambda) \cup \Segf{p}(\lambda) \cup \Segb{1+p}(T[1..p]) \,.
    \end{gather*}

    \paragraph*{Recurrence relations.}
    The algorithm recursively compute $L_{X_1X_2}[i_1,i_2,p_1,p_2]$ for $0 \le p_a \le g_a$ and $0<i_a \le n_a$.

    Suppose $T_1[i_1] \ne T_2[i_2]$.
    In this case, any common subsequence $u$ of $T_1[1..i_1]$ and $T_2[1..i_2]$ is a common subsequence of $T_1[1..i_1-1]$ and $T_2[1..i_2]$ or $T_1[1..i_1]$ and $T_2[1..i_2-1]$.
    Every subsequence $u$ of $T[1..i-1]$ is that of $T[1..i]$, where it may be classified with different parameters.
    Notice that
    \begin{itemize}
      \item if $u \in \SegSub{h}(T[1..i-1])$ or $u \in \SegSuf{h}(T[1..i-1])$, \\ then $u \in \SegSub{h}(T[1..i])$ and $u \in \SegSuf{h+1}(T[1..i])$;
      \item if $u \in \Segb{\ell}(T[1..i-1])$, then $u \in \Segb{\ell+1}(T[1..i])$;
      \item if $u \in \Segf{\ell}(T[1..i-1])$, then $u \in \Segb{\ell}(T[1..i])$.
    \end{itemize}
    The converse of the above observation holds provided that the last character $T[i]$ of $T[1..i]$ is not used in $u$.
    Say, $u \in \Segb{\ell}(T[1..i])$ implies $u \in \Segb{\ell-1}(T[1..i-1])$ or $u \in \Segf{\ell}(T[1..i-1])$ for $\ell \ge 1$.
    On the other hand, the definition of $\Segf{\ell}(T[1..i])$ requires to use the last character $T[i]$ in its member subsequences.
    Based on this argument, we obtain
    \begin{align*}
      L_{X_1X_2}[i_1,i_2,p_1,p_2] =
      \max( & \{\, L_{X_1Y_2}[i_1,i_2-1,p_1,p_2-q] \mid (Y_2,q) \in \Phi(X_2,p_2) \}
      \\ {}\cup{} & \{\, L_{Y_1X_2}[i_1-1,i_2,p_1-q,p_2] \mid (Y_1,q) \in \Phi(X_1,p_1) \} )
    \end{align*}
    where 
    \begin{align*}
      \Phi(\mrm{B},p) &= \{ (\mrm{B},p),(\mrm{F},p) \},
      \quad\quad
      \Phi(\mrm{F},p) = \begin{cases} \{ (\mrm{B},p-1),(\mrm{F},p-1) \} &\text{if $p > 0$},
        \\  \emptyset &\text{if $p=0$,}
      \end{cases}
      \\ 
        \Phi(\mrm{\ol{B}},p) &=\begin{cases}
          \{ (\mrm{\ol{B}},p-1),(\mrm{\ol{F}},p) \} &\text{if $p > 0$},
          \\    \{ (\mrm{\ol{B}},p),(\mrm{\ol{F}},p) \} &\text{if $p=0$,}
        \end{cases}
        \quad\quad
        \Phi(\mrm{\ol{F}},p) = \emptyset.
    \end{align*}

    Suppose $T_1[i_1] = T_2[i_2]$.
    In this case, a common subsequence $u$ of $T_1[1..i_1]$ and $T_2[1..i_2]$ may use the last characters of $T_1[1..i_1]$ and $T_2[1..i_2]$, and $u[1..|u|-1]$ is a common subsequence of $T_1[1..i_1-1]$ and $T_2[1..i_2-1]$.

    Observe that for $u = u'T[i]$,
    \begin{itemize}
      \item if $u' \in \SegSub{h}(T[1..i-1])$,\\ then $u \in \SegSub{h+1}(T[1..i])$ and $u \in \SegSuf{h+1}(T[1..i])$;
      \item if $u' \in \SegSuf{h}(T[1..i-1])$,\\ then $u \in \SegSub{h}(T[1..i])$ and $u \in \SegSuf{h}(T[1..i])$;
      \item if $u' \in \Segb{\ell}(T[1..i-1])$, then $u \in \Segf{\ell}(T[1..i])$;
      \item if $u' \in \Segf{\ell}(T[1..i-1])$, then $u \in \Segf{\ell+1}(T[1..i])$.
    \end{itemize}
    Since the longest common subsequence does not necessarily use the last matching characters of $T_1[1..i_1]$ and $T_2[1..i_2]$, we obtain
    \begin{align*}
      L_{X_1X_2}[i_1,i_2,p_1,p_2] =
      \max\big( & \{\, L_{X_1Y_2}[i_1,i_2-1,p_1,p_2-q] \mid (Y_2,q) \in \Phi(X_2,p_2) \}
        \\ {}\cup{} & \{\, L_{Y_1X_2}[i_1-1,i_2,p_1-q,p_2] \mid (Y_1,q) \in \Phi(X_1,p_1) \}
      \\ {}\cup{} & \{\, 1+ L_{Y_1Y_2}[i_1-1,i_2-1,p_1-q_1,p_2-q_2] \mid \\ & \quad\quad (Y_1,q_1) \in \Psi(X_1,p_1) \text{ and } (Y_2,q_2) \in \Psi(X_2,p_2) \} \big)
    \end{align*}
    where 
    \begin{align*}
      \Psi(\mrm{B},p) & = \begin{cases}
        \{ (\mrm{B},p-1),(\mrm{F},p-1) \} & \text{if $p>0$},
        \\ \emptyset & \text{if $p=0$},
      \end{cases}
      \\
      \Psi(\mrm{F},p) &= \begin{cases} \{ (\mrm{B},p-1),(\mrm{F},p) \} &\text{if $p >0$},
        \\ \emptyset &\text{if $p=0$},
      \end{cases}
      \\ 
      \Psi(\mrm{\ol{B}},p) &= \emptyset,
      \\
      \Psi(\mrm{\ol{F}},p) &=\begin{cases}
        \{ (\mrm{\ol{B}},p),(\mrm{\ol{F}},p-1) \} &\text{if $p >0$},
        \\    \{ (\mrm{\ol{B}},p),(\mrm{\ol{F}},p) \} &\text{if $p =0$.}
      \end{cases}
    \end{align*}

    It will be only tedious work to confirm that the above recurrence equations indeed maintains the table in accordance with Equation~(\ref{eq:indseg_recurrence}).

    \end{document}